\documentclass[12pt]{article}

\usepackage[T2A]{fontenc}
\usepackage[utf8]{inputenc}
\usepackage[english]{babel}
\usepackage{amsmath}
\usepackage{amsfonts}
\usepackage{mathrsfs}
\usepackage{amssymb}
\usepackage{bm}
\usepackage{cite}

\def\theequation{\arabic{section}.\arabic{equation}}
\numberwithin{equation}{section}

\newcommand{\be}{\begin{equation}}
\newcommand{\ee}{\end{equation}}
\newcommand{\bea}{\begin{eqnarray}}
\newcommand{\eea}{\end{eqnarray}}

\newcommand{\p}[1]{(\ref{#1})}

\begin{document}

\begin{titlepage}

\vspace*{0.7cm}

\begin{center}

{\LARGE\bf Massless finite and infinite spin representations}

\vspace{0.4cm}

{\LARGE\bf of Poincar\'{e} group in six dimensions}

\vspace{1.5cm}

{\large\bf I.L.\,Buchbinder$^{1,2}$\!\!,\ \ \  S.A.\,Fedoruk$^3$\!\!,\ \ \
A.P.\,Isaev$^{3,4}$\!\!,\ \ \ M.A.\,Podoinitsyn$^{3}$}

\vspace{1.5cm}

\ $^1${\it Department of Theoretical Physics,
Tomsk State Pedagogical University, \\
634041 Tomsk, Russia}, \\
{\tt joseph@tspu.edu.ru}

\vskip 0.5cm

\ $^2${\it National Research Tomsk State  University,}\\{\em 634050 Tomsk, Russia}

\vskip 0.5cm

\ $^3${\it Bogoliubov Laboratory of Theoretical Physics,
Joint Institute for Nuclear Research, \\
141980 Dubna, Moscow Region, Russia}, \\
{\tt fedoruk@theor.jinr.ru, isaevap@theor.jinr.ru, mpod@theor.jinr.ru}

\vskip 0.5cm

\ $^4${\it Physics Faculty, M.V.~Lomonosov Moscow State University, \\
119991 Moscow, Russia}, \\

\end{center}

\vspace{1.5cm}

\begin{abstract}

We study the massless irreducible representations of the Poincar\'e group in the six-dimensional Minkowski space. The Casimir operators are constructed and their eigenvalues are found. It is shown that the finite spin (helicity) representation is defined by two integer or half-integer numbers while the infinite spin representation is defined by the real parameter $\mu^2$ and one integer or half-integer number.

\end{abstract}

\vspace{3cm}

\noindent PACS: 11.10.Kk, 11.30.Cp, 11.30.-j

\smallskip
\noindent Keywords:  massless representations, helicity, infinite spin representations, Casimir operators \\

\vspace{1cm}

\end{titlepage}

\setcounter{footnote}{0}
\setcounter{equation}{0}

\newpage

\setcounter{equation}0
\section{Introduction}

Study of the various aspects of  field theory in higher dimensions
attracts much attention due to the remarkable and sometimes even
unexpected properties at classical and quantum levels. Many of such
properties are closely related to superstring theory which may be
treated as a theory of infinite number of higher spin fields in
higher dimensional space-time (see e.g. \cite{GSW}). In particular,
the low-energy limits of superstring theory are supersymmetric gauge
theories and supergravity in ten dimensions that after reduction
yield to field models in dimensions from ten to four. Since the
details of field theories are essentially defined by the space-time
symmetry, it seems useful to focus an attention on studying the
diverse specific properties of symmetry groups in the higher
dimensions.

The fundamental space-time background in relativistic theory is
Minkowski space where the basic symmetry is described by
Poincar\'{e} group. Theory of unitary irreducible representations of
Poincar\'{e} group in four dimensions was constructed in the pioneer
papers \cite{Wigner39,Wigner47,BargWigner}. The aspects of
unitary irreducible representations in higher dimensions and their
applications
are
considered in the papers \cite{BKRX}, \cite{KR},
\cite{BekBoul03}, \cite{BekBoul07}, \cite{BBAST} and in lectures
\cite{BekBoul} (see also the recent paper \cite{Wein}). Although the
generic scheme of constructing the representations of the
Poincar\'{e} group in any dimension seems can be realized on the
base of known method of induced representations (see e.g. \cite{BR},
\cite{IR}), many specific aspects important for classical and
quantum field theory deserve a separate attention and require
independent study. Some of such aspects are appropriate only for
each concrete dimension and can not be formulated at once for all
dimensions. For example, the spinor representations of the Lie
algebra of multidimensional Lorenz group are defined independently
for each space-time dimension. Therefore one can expect
that the structure of relativistic symmetry representations in higher
dimensions is much reacher and more complicated then in the
four-dimensional Minkowski space.

In this letter we construct the massless finite and infinite spin
irreducible representations of the Lie algebra of the Poincar\'{e} group in
six-dimensional Minkowski space. Some aspects of such
representations are considered in papers \cite{MezRTown14},
\cite{AMezTown} however many issues, especially the infinite spin
representations, were not addressed and complete analysis was not
done. Recently there was the paper \cite{KuzP} where the unitary
irreducible massless representations of the Poincar\'{e} group in
five-dimensional Minkowski space were constructed and some issues
related to representations in arbitrary dimensions were briefly
studied and the representations of super Poincar\'{e} group were
considered. The infinite spin representations were not addressed.

The letter is organized as follows. Section 2 is devoted to Casimir
operators and their properties in the six-dimensional standard
massless momentum reference frame. In section 3 we describe the
massless finite spin irreducible representations and show that they
are described by two integer or half-integer numbers. Section 4 is
devoted to infinite spin representations which are described by
arbitrary real parameter and a single integer or half-integer
number. Section 5 is a summary of the results.

\setcounter{equation}0
\section{Poincar\'{e} algebra and light-cone reference frame}

The generators $P_{m}$ and $M_{mn}=-M_{nm}$ of the Lie algebra $\mathfrak{iso}(1,D-1)$ of the Poincar\'{e} group in $D$-dimensional
space-time have the commutators
\begin{equation}
\label{P-M}
[P_{n},P_{k}]= 0 \; , \;\;\;\;\;
[M_{mn},P_{k}]=i\left(\eta_{mk}P_{n}-\eta_{nk}P_{m}\right)\,,
\end{equation}
\begin{equation}
\label{M-M}
[M_{mn},M_{kl}]=i\left(\eta_{mk}M_{nl}+\eta_{nl}M_{mk}-\eta_{ml}M_{nk}-\eta_{nk}M_{ml}\right)\,,
\end{equation}
where the $D$-vector indices run the values $m,n=0,1,\ldots,D-1$ and
we use the space-time metric $\eta^{mn}={\rm diag}(+1,\underbrace{-1,\ldots,-1}_{D-1})$.
We call the Lie algebra $\mathfrak{iso}(1,D-1)$ of the Poincar\'{e} group
as $D$-dimensional Poincar\'{e} algebra.

\subsection{Casimir operators of 6-dimensional Poincar\'{e} algebra}

We introduce the third rank tensor $W_{mnk}$ and the vector
$\Upsilon_m$ as the elements of the enveloping algebra
 of $\mathfrak{iso}(1,5)$  \cite{BKRX,MezRTown14}
\begin{eqnarray}
\label{W3} W_{mnk}&=& \varepsilon_{mnklpr} P^{l} M^{pr}\,, \\ [7pt]
\Upsilon_m &=& \varepsilon_{mnklpr} P^{n} M^{kl} M^{pr}\, .
\label{J5}
\end{eqnarray}
Here we use the totally antisymmetric tensor $\varepsilon_{mnklpr}$ and normalize it as
$\varepsilon_{012345}=1$. The operators \p{W3} and \p{J5}
satisfy the equations
\begin{eqnarray}
\label{W3-id}
&P^m W_{mnk}= 0\,, \quad & [P_l,W_{mnk}]= 0\,,\\ [8pt]
&P^m\Upsilon_m = 0\,, \quad & [P_l,\Upsilon_m ]= 0\,.\label{J5-id}
\end{eqnarray}
By using of these equations one can check that the operators
\begin{eqnarray}
\label{2-Cas-f}
C_2&:=&P^{m}P_{m} \,, \\ [7pt]
\label{4-Cas-f}
C_4&:=&\frac{1}{24}\,W^{mnk}W_{mnk}   \,, \\ [7pt]
C_6&:=&\frac{1}{64}\,\Upsilon^{m}\Upsilon_{m}
\label{J5-2-Cas-f}
\end{eqnarray}
are the  Casimir operators of the Poincar\'{e} algebra
$\mathfrak{iso}(1,5)$. It it clear
that $C_2$, $C_4$ and $C_6$ are second, fourth and sixth order
operators in the Poincar\'{e} algebra generators,
respectively.
The six-dimensional Poincar\'{e} algebra has no other additional Casimir operators (see comments in the Appendix).

Taking into account the expressions \p{W3}, \p{J5} we obtain
explicit form of the Casimir operators \p{2-Cas-f}, \p{4-Cas-f}, \p{J5-2-Cas-f}:
\begin{eqnarray}
\label{P-2-Cas}
C_2&=&P^{m}P_{m} \,, \\ [7pt]
\label{W3-2-Cas}
C_4&=&  \Pi^{m} \Pi_{m} \ - \ \frac12\,M^{mn}M_{mn} \, C_2  \,, \\ [7pt]
\nonumber
C_6&=&
-\, \Pi^k M_{km}\, \Pi_l M^{lm}
\ + \ \frac12\,\Big(M^{mn}M_{mn}-8\Big)\, C_4 \\ [7pt]
\label{J5-2-Cas}
&&
+\ \frac18\, \Big[M^{kl}M_{kl}\Big(M^{mn}M_{mn}-8\Big) +2M^{mn}M_{nk} M^{kl}M_{lm}\Big]\, C_2
\,,
\end{eqnarray}
where we introduce new vector $\Pi$ with components
\begin{equation}
\label{Pi-def}
\Pi_m := P^{k} \, M_{km} = M_{km}\, P^{k} - 5 i \, P_m \; ,
\end{equation}
which satisfy commutation relations (cf. (\ref{P-M}))
\begin{equation}
\label{pi-alg}
 [\Pi_n, \; \Pi_k] = -i \, M_{nk} \, C_2 \, , \;\;\;\;
 [M_{mn},\Pi_{k}]=i\left(\eta_{mk}\Pi_{n}-\eta_{nk}\Pi_{m}\right)\, .
\end{equation}

Further in this paper we consider the massless unitary representations of the algebra $\mathfrak{iso}(1,5)$
when the quadratic Casimir operator \p{P-2-Cas} is fixed as following:
\begin{equation}
\label{P2-0}
C_2 \equiv P^2=P^{m}P_{m}=0\,.
\end{equation}

\subsection{Standard massless momentum reference frame}

Let the algebra (2.1), (2.2) acts in the representation space
${\cal H}$ with basis vectors $|k, \sigma \rangle$,
where $\sigma$ is a set of eigenvalues of all operators
commuting with $P_m$ and
$P_m |k, \sigma \rangle = k_m |k, \sigma \rangle$.
We take the light-cone reference frame for massless
particle momentum $k^m = (k^0,k^a,k^5)=(k,0,0,0,0,k)$ in which momentum
operator \p{P2-0} has the standard form
\begin{equation}
\label{P-st}
P^0=P^5=k\,,\qquad P^a=0\,,\ \ \ a=1,2,3,4\,.
\end{equation}
We stress that all operator formulas presented in this Section
(and written in the light-cone frame) should be understood
as a result of their action on the
subspace ${\cal H}_k \subset {\cal H}$ spanned by vectors
$|k, \sigma \rangle$ with fixed
light-cone momentum $k_m$.

The transition to this light-cone reference frame is conveniently performed in the light-cone basis where
any $6D$ vector $X^m=(X^0,X^a,X^5)$ has the light-cone coordinates
$X^m=(X^+,X^-,X^a)$, where
\begin{equation}
\label{X-lc}
X^\pm=\frac{1}{\sqrt{2}}\left(X^0\pm X^5\right)\,,\qquad X_\pm=\frac{1}{\sqrt{2}}\left(X_0\pm X_5\right) \qquad
\Rightarrow \qquad  X^\pm= X_\mp \; .
\end{equation}
Then, in the light-cone basis the contraction
of two $6D$ vectors $X^m$ and $Y^m$ is
\begin{equation}
\label{XY-lc}
\begin{array}{rcl}
X^m Y_m&=&X^+ Y_+ + X^- Y_- + X^a Y_a  \\ [0.2cm]
&=& \eta^{-+} \, X_- Y_+  +   \eta^{+-}\, X_+ Y_- + \eta^{ab}\,  X_b Y_a \; =
 \; X_- Y_+  +   X_+ Y_-   - X_a Y_a \,,
\end{array}
\end{equation}
where we use the light-cone metric
$\eta^{\pm \mp}=\eta_{\pm \mp}=1$, $\eta^{\pm \pm}=\eta_{\pm \pm}=0$, $\eta^{ab}=\eta_{ab}=-\delta_{ab}$.
In the light-cone basis the total antisymmetric tensor $\varepsilon_{mnklpr}$ has components
$$
\varepsilon_{-+abcd}=-\varepsilon_{+-abcd}=\varepsilon^{+-abcd}
=-\varepsilon^{-+abcd}=\varepsilon_{abcd} \; ,
$$
and we normalize the
antisymmetric tensors $\varepsilon_{mnk\ell pr}$ and $\varepsilon_{abcd}$
 as $\epsilon_{012345}=1$ and $\epsilon_{1234}=1$.

In the light-cone basis the standard momentum \p{P-st} has the components
\begin{equation}
\label{P-st1}
P^+=P_-=\sqrt{2}k\,,\qquad P^-=P_+=0\,,\qquad P^a=0\,,\ \ \ a=1,2,3,4\,.
\end{equation}
Thus, in the light-cone reference frame \p{P-st1}
the Casimir operators \p{W3-2-Cas}, \p{J5-2-Cas} take the form\,\footnote{When we deduce \p{W2-st} and \p{J5-st} it is necessary,
since we project all operator relations to the subspace ${\cal H}_k$,
first move all operators $P_m$ in the expressions
\p{W3-2-Cas} and \p{J5-2-Cas}
to the right and only then perform the substitution \p{P-st1}.}
\begin{eqnarray}
\label{W2-st}
\hat C_4&=&-\hat\Pi_a\hat\Pi_a \,, \\ [7pt]
\label{J5-st}
\hat C_6&=& \hat\Pi_b M_{ba} \,\hat\Pi_c M_{ca} \ - \
\frac12\, M_{bc}M_{bc}\,\hat\Pi_a\hat\Pi_a \,,
\end{eqnarray}
where we introduce Hermitian operators
\begin{equation}
\label{calPJ-def}
\hat\Pi_a:=\sqrt{2}kM_{+a}\,.
\end{equation}
Formula (\ref{W2-st}) directly follows from \p{W3-2-Cas},
while derivation of (\ref{J5-st}) from \p{J5-2-Cas} takes some efforts.

In view of \p{M-M}
the operators $\hat\Pi_a$ \p{calPJ-def} and $M_{ab}$,
which generate \p{W2-st} and \p{J5-st}, form the Lie
algebra of ${ISO}(4)$ group
\begin{equation}
\label{P-J}
[\hat\Pi_{a},\hat\Pi_{b}]=0\,,\qquad
[\hat\Pi_{a},M_{bc}]=i\left(\delta_{ab}\hat\Pi_{c}
-\delta_{ac}\hat\Pi_{b}\right)\,,
\end{equation}
\begin{equation}
\label{J-J}
[M_{ab},M_{cd}]=
i\left(\delta_{bc}M_{ad}-\delta_{bd}M_{ac}
+\delta_{ac}M_{db}-\delta_{ad}M_{cb}\right) \; ,
\end{equation}
and therefore generate the isometries of the four-dimensional Euclidean space.
As a result, the operators $\hat C_4$ and $\hat C_6$ defined in \p{W2-st} and \p{J5-st} are the Casimir operators
of the $\mathfrak{iso}(4)$ algebra.

Six generators of rotations $M_{ab}$ in four-dimensional Euclidean space are decomposed into the sum
\begin{equation}
\label{M-Mpm}
M_{ab}=M_{ab}^{(+)} + M_{ab}^{(-)}\,,
\end{equation}
where
\begin{equation}
\label{Mpm}
M^{(\pm)}_{ab} := \frac12 \left(M_{ab}\pm\frac12\,\epsilon_{abcd}M_{cd}\right)
\end{equation}
are (anti)selfdual parts. They satisfy the identities
\begin{equation}
\label{Mpm-ed}
M^{(\pm)}_{ab} = \pm\frac12\,\epsilon_{abcd}M^{(\pm)}_{cd}\,.
\end{equation}
The generators \p{Mpm} form the algebra
\begin{equation}
\label{M-M-pm}
[M^{(\pm)}_{ab},M^{(\pm)}_{cd}]=
i\Big(\delta_{bc}M^{(\pm)}_{ad}-\delta_{bd}M^{(\pm)}_{ac}+\delta_{ac}M^{(\pm)}_{db}-\delta_{ad}M^{(\pm)}_{cb}\Big)
\,,\qquad [M^{(+)}_{ab},M^{(-)}_{cd}]=0\,,
\end{equation}
which is direct sum of two algebras with three generators $M^{(+)}_{ab}$ and with three generators $M^{(-)}_{ab}$ respectively.
Each of these algebras, containing three generators $M^{(+)}_{ab}$ or $M^{(-)}_{ab}$, is the $\mathfrak{su}(2)$ algebra.

This becomes clear (see e.g. \cite{IR})
after using the ‘t Hooft symbols \cite{tHooft}.
The ‘t\,Hooft symbols $\eta^{\mathrm{i}}_{ab}=-\eta^{\mathrm{i}}_{ba}$,
$\mathrm{i}=1,2,3$ and $\bar\eta^{\mathrm{i}^\prime}_{ab}=-\bar\eta^{\mathrm{i}^\prime}_{ba}$, $\mathrm{i}=1,2,3$
are (anti-)selfdual tensors with respect to the $SO(4)$ indices $a,b$:
\begin{equation}
\label{eta-sd}
\eta^{\mathrm{i}}_{ab} = \frac12\,\epsilon_{abcd}\eta^{\mathrm{i}}_{cd}\,,\qquad
\bar\eta^{\mathrm{i}^\prime}_{ab} = -\frac12\,\epsilon_{abcd}\bar\eta^{\mathrm{i}^\prime}_{cd}\,.
\end{equation}
Below we use the following standard representations
for the ‘t Hooft symbols
\begin{equation}
\label{eta}
\eta^{\mathrm{i}}_{ab}=
\left\{
\begin{array}{l}
\epsilon_{\mathrm{i}ab} \qquad a,b=1,2,3,\\
\delta_{\mathrm{i}a} \qquad b=4,
\end{array}
\right.
\qquad\quad
\bar\eta^{\mathrm{i}^\prime}_{ab}=
\left\{
\begin{array}{l}
\epsilon_{\mathrm{i}^\prime ab} \qquad a,b=1,2,3,\\
-\delta_{\mathrm{i}^\prime a} \qquad b=4.
\end{array}
\right.
\end{equation}
Due to the properties \p{eta-sd} the ‘t\,Hooft symbols connect (anti-)selfdual $SO(4)$ tensors $M^{(\pm)}_{ab}$ \p{Mpm} with
the $SO(3)$ vectors $M^{(+)}_{\mathrm{\,i}}$, $M^{(-)}_{\mathrm{\,i}^\prime}$ by means of the following relations
\begin{equation}
\label{Mpm-eta}
M^{(+)}_{ab} = -\eta^{\mathrm{i}}_{ab}M^{(+)}_{\mathrm{\,i}}\,,\qquad
M^{(-)}_{ab} = -\bar\eta^{\mathrm{i}^\prime}_{ab}M^{(-)}_{\mathrm{\,i}^\prime}\,.
\end{equation}
Such defined operators $M^{(+)}_{\mathrm{\,i}}$ and $M^{(-)}_{\mathrm{\,i}^\prime}$ form two $\mathfrak{su}(2)$ algebras with standard form of the commutators
\begin{equation}
\label{M-M-2}
[M^{(+)}_{\mathrm{\,i}},M^{(+)}_{\mathrm{\,j}}]=i\epsilon_{\mathrm{i}\mathrm{j}\mathrm{k}}M^{(+)}_{\mathrm{\,k}}\,,\qquad
[M^{(-)}_{\mathrm{\,i}^\prime},M^{(-)}_{\mathrm{\,j}^\prime}]=
i\epsilon_{\mathrm{i}^\prime\mathrm{j}^\prime\mathrm{k}^\prime}M^{(-)}_{\mathrm{\,k}^\prime}\,,\qquad
[M^{(+)}_{\mathrm{\,i}},M^{(-)}_{\mathrm{\,j}^\prime}]=0\,.
\end{equation}

In term of the operators \p{Mpm-eta} the Casimir \p{J5-st} takes the form
(we use the equalities
$\eta^{\mathrm{i}}_{ab}\eta^{\mathrm{j}}_{ab}=4\delta^{\mathrm{i}\mathrm{j}}$,
$\bar\eta^{\mathrm{i}^\prime}_{ab}\bar\eta^{\mathrm{j}^\prime}_{ab}=4\delta^{\mathrm{i}^\prime\mathrm{j}^\prime}$ and
$\eta^{\mathrm{i}}_{ab}\bar\eta^{\mathrm{j}^\prime}_{ab}=0$)
\begin{equation}
\label{J5-st-a}
\hat C_6=2M^{(+)}_{\mathrm{\,i}} M^{(-)}_{\mathrm{\,j}^\prime} \,
\eta^{\mathrm{i}}_{ab} \bar\eta^{\mathrm{j}^\prime}_{ac} \, \hat\Pi_b \hat\Pi_c
- \left( M^{(+)}_{\mathrm{\,i}}M^{(+)}_{\mathrm{\,i}}+
M^{(-)}_{\mathrm{\,i}^\prime} M^{(-)}_{\mathrm{\,i}^\prime}\right)\hat\Pi_a \hat\Pi_a\,.
\end{equation}

Thus, in massless case \p{P2-0} unitary irreducible representations are defined by the eigenvalues
of the $\mathfrak{iso}(4)$ Casimir operators \p{W2-st} and \p{J5-st}.
In case of this noncompact symmetry there are two different cases defined the value of Casimir operator \p{W2-st},
i.e. square of ``four-translation'' generator $\hat\Pi_a$. So, in next sections we consider following unitary massless representations:
\begin{itemize}
\item
\textbf{Finite spin (helicity)  representations}. \\
In these cases
the ${SO}(4)$ four-vector $\hat\Pi_a$ has zero norm:
\begin{equation}
\label{P-2-0}
\hat\Pi_a\hat\Pi_a=0 \,.
\end{equation}
\item
\textbf{Infinite (continuous) spin representations}. \\
In case of these representations the
Euclidean four-vector $\hat\Pi_a$ has nonzero norm:
\begin{equation}
\label{P-2-n0}
\hat\Pi_a\hat\Pi_a=\mu^2\neq 0 \,.
\end{equation}
\end{itemize}

In next sections we consider these massless representations in details.

\setcounter{equation}0
\section{Massless finite spin representations}

This case is characterized by the fulfillment of condition \p{P-2-0},
which implies that all components $\hat\Pi_a$
(since they are Hermitian operators) of the Euclidean vector are zero:
\begin{equation}
\label{P-2a-0}
\hat\Pi_a=0 \qquad \mbox{at all} \qquad a=1,2,3,4 \,.
\end{equation}
As result, the Casimir operators
\p{W2-st} and \p{J5-st} are vanish in this case: $\hat C_4=0$ and $\hat C_6=0$.
In passing from this light-cone reference frame to an arbitrary frame, we get that
all Casimir operators (\ref{W3-2-Cas}), (\ref{J5-2-Cas})
on the massless finite spin states take zero values
(see also \cite{MezRTown14}):
\begin{equation}
\label{Cas-0-0}
C_4=0,\qquad C_6=0 \, ,
\end{equation}
and in view of (\ref{P2-0}) we have $\Pi^k \Pi_k=0$
and $\Pi^k M_{km} \Pi_\ell M^{\ell m}=0$.

Due to \p{P-2a-0} the Euclidean four-translations are realized trivially in case of these representations.
As a result such representations
of  $ISO(4)$ are finite dimensional.
Each such  representation defines some $6D$ standard massless representation.
As we saw above, such representations are induced from irreducible $SO(4)$ representations.
Let us show below that the Casimir operators of the stability
subgroup $SO(4)$ define the $6D$ helicity operators.

\subsection{$6D$ helicity operators}

First, let us consider the vector $\Upsilon_m$ defined in \p{J5}.
In the case $C_6=0$, according to \p{J5-2-Cas-f},
we have $\Upsilon_m \Upsilon^m =0$ and,
in the light-cone reference frame \p{P-st}, \p{P-st1},
the components of $6D$ vector $\Upsilon$ are
\begin{equation}
\label{J-P-st-01}
\Upsilon^+=\Lambda_1 P^+\,,\qquad  \Upsilon^-=\Upsilon_a=0\,,
\end{equation}
where we have
\begin{equation}
\label{Lambda-st-01}
\Lambda_1 \ := \ \epsilon_{abcd}M_{ab}M_{cd}\,.
\end{equation}
This operator is the Casimir operator of the $\mathfrak{so}(4)$
algebra.

 The conditions \p{J-P-st-01}
 demonstrate that vectors $\Upsilon$ and $P$
 are collinear in the light-cone reference frame and this
 property is conserved in any reference frame.
Namely, the relations \p{J5-id} show that the light like vector
$\Upsilon$ is transverse to the vector $P$ and its components
$\Upsilon_m$ commute with $P_k$. Therefore,  the vector $\Upsilon_m$
is proportional to the vector $P_m$:
\begin{equation}
\label{J-P-0} \Upsilon_m=\Lambda_1 P_m.
\end{equation}
This relation was also pointed out in
\cite{MezRTown14,KuzP}. Note that the operator
\p{Lambda-st-01} can be represented in the form
\begin{equation}
\label{Lambda-01}
\Lambda_1 :=\frac{\Upsilon_0}{P_0}\, .
\end{equation}
This expression appears for the $4D$ helicity operator
 when $\Upsilon_m$ is replaced by $W_m$.
Due to the relations
\begin{equation}
\label{Lambda-M-alg}
[M_{0i},\Lambda_1]=\frac{i}{P_0}
\big(\Upsilon_i-\Lambda_1 P_i\big)=0 \, , \;\;\;\;
[M_{ik},\Lambda_1] =0 = [P_{k},\Lambda_1] \, , \;\;\;\;\; (i,k=1,\ldots,5) \, ,
\end{equation}
we conclude that the operator \p{Lambda-01} is invariant with respect to the $6D$ Poincare symmetry.
Therefore, the operator $\Lambda_1$, defined in \p{Lambda-01}, is
a $6D$ analog of the helicity operator and it coincides
with one of $\mathfrak{so}(4)$ Casimir
operators in the light-cone reference frame.

We note that irreducible $\mathfrak{so}(4)$ representations are characterized
by two quadratic Casimir operators.
The second Casimir operator arises as helicity operator if we use the construction proposed in \cite{KuzP}.
Indeed, by using the prescription of \cite{KuzP}, one can construct another (third order in generators
of $\mathfrak{iso}(1,5)$) vector with components\footnote{In the definition of
antisymmetrization of $n$ indices, we use the factor $n!$, i.e.
$$
A_{[m}B_nC_{k]}=\frac{1}{3!}\,\Big(A_{m}B_nC_{k}-A_{m}B_kC_{n}
+\mbox{cyclic permutations} \Big)\,.
$$}
\begin{equation}
\label{S-3-0}
S_m \ := \ 3 M^{nk}P_{[m}M_{nk]} \ = \ M^{nk} M_{nk}P_{m} - 2M^{kn} M_{mn}P_{k}\,.
\end{equation}
The square of this $6D$ vector
  is
\begin{equation}
\label{S3-id}
S^m S_m= M^{nm}M_{nm} M^{lk}M_{lk} P^2 +4 \Big[\Pi^k M_{km} \Pi_l M^{lm} - M^{nm}M_{nm}(\Pi^l\Pi_l +P^2) + \Pi^l \Pi_l\Big]\,,
\end{equation}
while its contraction with $6D$ vector momentum $P_m$ gives
\begin{equation}
\label{S3-com1}
P^mS_m =  M^{mn}M_{mn} P^2-2\Pi^m \Pi_m \equiv
- 2 \, C_4 \; ,
\end{equation}
and the commutators of $S_m$ and $P_n$ are
\begin{equation}
\label{S3-com2}
[S_m,\; P_n]=2i M_{mn}P^2+4i\Pi_{[m} P_{n]} \; .
\end{equation}

For the massless finite spin representations,
defined by the conditions \p{P2-0}, \p{P-2a-0} and  \p{Cas-0-0},
equations \p{S3-id}, \p{S3-com1} and \p{S3-com2} are
reduced to
\begin{equation}
\label{S3-id0}
S^m S_m= 0\,,  \qquad  P^m S_m = 0\,, \qquad  [S_m,P_n ]= 0\,,
\end{equation}
which are the same as conditions \p{J5-id}
for light-like vectors $\Upsilon$ and $P$.
So, in the case of massless finite spin representations,
the vectors $P_m$ and $ S_m$ are also proportional to each other.
One can check this in the light-cone reference frame,
when subject to the conditions \p{P-st1} and \p{P-2a-0} the components of the $6D$ vector \p{S-3-0} are equal to
\begin{equation}
\label{J-P-st-02}
S^+=\Lambda_2 P^+\,,\qquad  S^-=S_a=0\,,
\end{equation}
where the operator
\begin{equation}
\label{Lambda-st-02}
\Lambda_2 \ := \ M_{ab}M_{ab}
\end{equation}
is second $\mathfrak{so}(4)$ Casimir operator.

Due to the relations \p{S3-id0} in general frame the relations \p{J-P-st-02} take the form
\begin{equation}
\label{S-P-0}
S_m=\Lambda_2 P_m\,,
\end{equation}
where the operator $\Lambda_2$ \p{Lambda-st-02} defines second helicity operator and has equivalent ``covariant'' form
\begin{equation}
\label{Lambda-02}
\Lambda_2 :=\frac{S_0}{P_0}\,.
\end{equation}

So these massless representations of finite spin are characterized by the pair $(\lambda_1,\lambda_2)$,
where real numbers $\lambda_{1,2}$ define the eigenvalue of
the Casimir operators $\Lambda_{1,2}$ presented in \p{Lambda-01} and \p{Lambda-02}, respectively.

Using \p{M-Mpm} and \p{Mpm-eta} we represent helicity operators \p{Lambda-st-01} and \p{Lambda-st-02}
in the form
\begin{eqnarray}
\label{helic-1}
\Lambda_1&=& 2\left(M^{(+)}_{ab}M^{(+)}_{ab}- M^{(-)}_{ab}M^{(-)}_{ab}\right) \ =  \
8\left( M^{(+)}_{\mathrm{\,i}}M^{(+)}_{\mathrm{\,i}} - M^{(-)}_{\mathrm{\,i}^\prime}M^{(-)}_{\mathrm{\,i}^\prime}\right)\,, \\ [7pt]
\label{helic-2}
\Lambda_2&=& M^{(+)}_{ab}M^{(+)}_{ab}+ M^{(-)}_{ab}M^{(-)}_{ab} \ =  \
4\left( M^{(+)}_{\mathrm{\,i}}M^{(+)}_{\mathrm{\,i}} + M^{(-)}_{\mathrm{\,i}^\prime}M^{(-)}_{\mathrm{\,i}^\prime}\right) \,.
\end{eqnarray}
In case of unitary representations, the operators $M^{(+)}_{\mathrm{\,i}}M^{(+)}_{\mathrm{\,i}}$ and
$M^{(-)}_{\mathrm{\,i}^\prime}M^{(-)}_{\mathrm{\,i}^\prime}$ equal $j_+(j_+ +1)$ and $j_-(j_- +1)$ respectively.
Therefore, the eigenvalues of the helicity operators \p{Lambda-01} and \p{Lambda-02} take the values
\begin{eqnarray}
\label{helic-1-val}
\lambda_1&=&
8 j_+(j_+ +1) - 8 j_-(j_- +1) \,, \\ [7pt]
\label{helic-2-val}
\lambda_2&=&
4 j_+(j_+ +1) +4 j_-(j_- +1) \,,
\end{eqnarray}
where $j_\pm$ are integer or half-integer numbers in case of the unitary representations.

We note that the standard $4D$ helicity operator is invariant under
proper $SO(1,3)$ rotations but changes its sign under improper $O(1,3)$
rotations (reflections). We have the same property for $\Lambda_1$
but it is not the case for $\Lambda_2$.

\subsection{Examples}

Here we will demonstrate the use of the obtained formulas for determining the helicities
on the examples of some massless finite spin fields.
 To clarity and avoid technical complications, we will consider only bosonic integer-spin fields.

Since the irreducible massless representations of the $6D$ Poincar\'{e} group are induced by
the irreducible $SO(4)$ representations in the light-cone reference frame,
we will use the following procedure.

Below, in all examples of this section, we first consider a fixed irreducible $SO(4)$ representation and determine the values of the helicities.
Here we will use the defining representation for the $\mathfrak{so}(4)$ generators
\begin{equation}
\label{so4-vect-1}
(\mathcal{M}_{ab})_{eg} =i(\delta_{ae}\delta_{bg}-\delta_{ag}\delta_{be})\,.
\end{equation}
Then we reconstruct the corresponding $6D$ field, for which the equations of motion and gauge fixing show
that the independent components are exactly those $SO(4)$ fields
which were considered earlier in the Euclidean four-dimensional picture.

\subsubsection{Vector field}

Let us consider the $SO(4)$ vector field $A_a$.
In this case the $\mathfrak{so}(4)$ generators coincide with \p{so4-vect-1}:
\begin{equation}
\label{so4-vect-1-1}
(M_{ab})_{eg} =(\mathcal{M}_{ab})_{eg}\,.
\end{equation}
Then, the $SO(4)$ Casimir operators take the form
\begin{equation}
\label{hel-vect-1}
\begin{array}{rcl}
(\Lambda_1)_{eg}&=&\epsilon_{abcd}(M_{ab}M_{cd})_{eg} \ = \ 0\,, \\ [7pt]
(\Lambda_2)_{eg}&=&(M_{ab}M_{ab})_{eg} \ = \ 6 \delta_{eg}\,.
\end{array}
\end{equation}
When acting on the $SO(4)$ vector field $A_a$, the operators \p{hel-vect-1} give the following values of helicities:
\begin{equation}
\label{hel-vect}
\lambda_1=0\,,\quad \lambda_2=6\,;\qquad j_+=j_-=\frac12\,.
\end{equation}

This Euclidean vector field $A_a$ describes physical components of the $6D$ vector gauge field $A_m$.
In the momentum representation
 the $U(1)$ massless gauge field $A_m$ is
described by the equations of motion
\begin{equation}
\label{eq-6-vect}
P^m F_{mn} = 0\,,
\end{equation}
where $F_{mn}=i(P_mA_n-P_nA_m)$ is the field strength,
and determined up to gauge transformations
\begin{equation}
\label{g-6-vect}
\delta A_m = i P_m \varphi \; .
\end{equation}
One of the possible  gauge fixing
for transformations \p{g-6-vect} is the light-cone gauge (see e.g. \cite{Sieg})
\begin{equation}
\label{lc-6-vect}
A^+ = 0\,.
\end{equation}
Then in the light-cone frame \p{P-st1}, the equations of motion \p{eq-6-vect} give $A^- = 0$
and independent field is given by
the transverse part $A_a$ of the $6D$ gauge field $A_m$.

\subsubsection{Second rank symmetric tensor field}

Now we consider the $SO(4)$ second rank tensors. In this case the $\mathfrak{so}(4)$ generators take the matrix form
\begin{equation}
\label{so4-vect-2}
(M_{ab})_{e_1e_2,g_1g_2} = \bigl((\mathcal{M}_{ab})_1 +
(\mathcal{M}_{ab})_2\bigr)_{e_1e_2,g_1g_2} =(\mathcal{M}_{ab})_{e_1g_1}\delta_{e_2g_2}+\delta_{e_1g_1}(\mathcal{M}_{ab})_{e_2g_2}
\end{equation}
and the $SO(4)$ Casimir operators are
\begin{equation}
\label{hel-vect-2}
\begin{array}{rcl}
(\Lambda_1)_{e_1e_2,g_1g_2}&=&
\epsilon_{abcd}\bigl(M_{ab}M_{cd} \bigr)_{e_1e_2,g_1g_2} \, =
2 \, \epsilon_{abcd}\bigl(
(M_{ab})_1(M_{cd})_2 \bigr)_{e_1e_2,g_1g_2} \, =
\ 8\, \epsilon_{e_1e_2g_1g_2}\,, \\ [10pt]
(\Lambda_2)_{e_1e_2,g_1g_2}&=&(M_{ab}M_{ab})_{e_1e_2,g_1g_2}\, =\,
\bigl((M^2_{ab})_1 + (M^2_{ab})_2 + 2 (M_{ab})_1(M_{ab})_2
 \bigr)_{e_1e_2,g_1g_2} \, = \\ [5pt]
&=& 12 \; \delta_{e_1g_1}\delta_{e_2g_2} + 4( \delta_{e_1g_2}\delta_{e_2g_1}-\delta_{e_1e_2}\delta_{g_1g_2})\,.
\end{array}
\end{equation}

First, we consider the $SO(4)$ second rank tensor $\hat h_{ab}$, which is symmetric $\hat h_{ab}=\hat h_{ba}$
and traceless $\hat h_{aa}$.
On this field the helicity operators \p{hel-vect-2} take the values
\begin{equation}
\label{hel-gr}
\lambda_1=0\,,\quad \lambda_2=16\,;\qquad j_+=j_-=1\,.
\end{equation}
Let us show that this field $\hat h_{ab}$ describes the physical components of the $6D$ linearized gravitational field.

The $6D$ linearized gravitational field $h^{mn}=h^{nm}$ is determined by the well known equations of motion
\begin{equation}
\label{eq-6-gr}
P^2h^{mn}-P^m P_kh^{nk}-P^n P_kh^{mk}+P^m P^n h_{k}{}^{k} = 0\,,
\end{equation}
and has gauge invariance
\begin{equation}
\label{g-6-gr}
\delta h^{mn} = i P^{(m} \varphi^{n)} \; .
\end{equation}
For the transformations \p{g-6-gr} we can put again the light-cone gauge (see also \cite{Sieg})
\begin{equation}
\label{lc-6-gr}
h^{+m} = 0\,.
\end{equation}
The equations of motion \p{eq-6-gr} produce $h^{-m} = 0$, $h_{a}{}^a = 0$ in the light-cone frame \p{P-st1}.
As a result, nonvanishing physical components
of the $6D$ gravity field $h_{mn}$ are given
by the traceless part $\hat h_{ab}$ of its transverse components $h_{ab}$.

\subsubsection{Third rank (anti-)selfdual antisymmetric tensor fields\label{3nrt}}

Now we consider the $SO(4)$ antisymmetric tensors of the second rank $B^{(\pm)}_{ab}=-B^{(\pm)}_{ba}$, which are (anti-)selfdual
\begin{equation}
\label{sd-4-tens}
B^{(\pm)}_{ab} = \pm\frac12\,\epsilon_{abcd}B^{(\pm)}_{cd}\,.
\end{equation}
These tensors form the spaces of two $SO(4)$ irreducible representations
which make up the $SO(4)$ reducible
representation in the space of all antisymmetric rank 2 tensors
associated to Young diagram $[1^2] \equiv {\scriptsize \begin{array}{|c|}
\hline
\;\, \\ [0.15cm] \hline
\;\,  \\ [0.15cm]
\hline
\end{array}
}$\,. In this case
the $\mathfrak{so}(4)$ generators $M_{ab}$ and helicity operators
 $\Lambda_1,\Lambda_2$ have the same expressions
\p{so4-vect-2} and \p{hel-vect-2}.
 Then the eigenvalues of the operators $\Lambda_1,\Lambda_2$
 and $(M_{\mathrm{\,i}}^{(\pm)}M_{\mathrm{\,i}}^{(\pm)})$ are given by numbers
\begin{equation}
\label{hel-tens1}
\lambda_1=16\,,\quad \lambda_2=8\,;\qquad j_+=1\,,\quad j_-=0
\end{equation}
on the space of the selfdual fields $B^{(+)}_{ab}$, and by
\begin{equation}
\label{hel-tens2}
\lambda_1=-16\,,\quad \lambda_2=8\,;\qquad j_+=0\,,\quad j_-=1
\end{equation}
on the space of the anti-selfdual fields $B^{(-)}_{ab}$.
It is clear that these $SO(4)$ (anti-)selfdual fields $B^{(\pm)}_{[ab]}$
are independent components of the $6D$ massless (anti-)selfdual
3-rank fields $B^{(\pm)}_{mnk}$  which satisfy the identities
\begin{equation}
\label{sd-6-tens}
B^{(\pm)}_{mnk}= \pm \frac{1}{3!}\, \varepsilon_{mnklpr} B^{(\pm)}{}^{lpr}\,.
\end{equation}
So, the equations of motion of the $6D$ massless fields $B^{(\pm)}_{mnk}$ are
\begin{equation}
\label{eq-6-tens}
\mbox{a)} \ \ P^m B^{(\pm)}_{mnk} = 0\,,\qquad\quad \mbox{b)} \ \ P_{[m} B^{(\pm)}_{nkl]}=0\,, \qquad\quad \mbox{c)} \ \ P^2 B^{(\pm)}_{nkl}=0 \,.
\end{equation}
Then in the light-cone frame \p{P-st1} the equations (\ref{eq-6-tens}a) give $B^{(\pm)}{}^{-mn}=0$
whereas the equations (\ref{eq-6-tens}b) produce $B^{(\pm)}{}^{abc}=0$.
As a result, independent fields of the $6D$
 tensors $B^{(\pm)}_{mnk}$ are
the $SO(4)$ (anti-)selfdual fields $B^{(\pm)}{}^{+ab}\equiv B^{(\pm)}{}^{ab}$
which are subjected the $SO(4)$ (anti-)selfdual conditions \p{sd-4-tens}
due to the $6D$ (anti-)selfdual conditions \p{sd-6-tens}.

\vspace{0.2cm}

\noindent
{\bf Remark.}  One can generalize this example to the case of
special $3n$-rank selfdual and anti-selfdual 6-dimensional
tensor fields. These fields
correspond to $SO(4)$ irreducible representations
in spaces of $2n$-rank traceless selfdual
and anti-selfdual tensors
 with components $B^{(\pm)}_{a_1...a_{2n}}$ be symmetrized
 in accordance to the Young diagram
 $[n^2] \equiv {\scriptsize \begin{array}{|c|c|c|c|}
\hline
\;\, & \;\, & \! \dots \! & \;\,  \\ [0.15cm] \hline
\; &  \; & \! \dots \! & \; \\ [0.15cm]
\hline
\end{array}
}$ . It is clear that for highest weights of such selfdual and anti-selfdual representations
of $SO(4)$ we have respectively $j_+=n,j_-=0$ and $j_+=0,j_-=n$
and in view of (\ref{helic-1-val}) and (\ref{helic-2-val})
we obtain the eigenvalues
of helicity operators
$\lambda_1= 8 n (n+1)$, $\lambda_2 = 4 n (n+1)$
and $\lambda_1= - 8 n (n+1)$, $\lambda_2 = 4 n (n+1)$
which is a generalization of (\ref{hel-tens1})
and (\ref{hel-tens2}).

\setcounter{equation}0
\section{Massless infinite (continuous) spin representations}

In this case, when the condition \p{P-2-n0} is satisfied and the Euclidean four-vector $\hat\Pi_a$ is nonzero.
Then here the representations of the $ISO(4)$ group, which induce the $6D$ relativistic massless representations,
are infinite dimensional.

In case of these representations the Casimir operator \p{W2-st} has nonvanishing eigenvalue
\begin{equation}
\label{W-2-n0}
C_4 \ = \ \hat C_4 \ = \ -\mu^2 \,, \qquad \mu\neq 0\,.
\end{equation}
Moreover, for the orbits \p{P-2-n0} we can take the basis with with nonzero only the fourth component:
\begin{equation}
\label{Pi-2-n0}
\hat\Pi_1=\hat\Pi_2=\hat\Pi_3=0\,,\qquad
\hat\Pi_4=\mu \,.
\end{equation}
Then taking into account $\eta^{\mathrm{i}}_{a4}=\delta_{\mathrm{i} a}$ and $\bar\eta^{\mathrm{i}^\prime}_{a4}=-\delta_{\mathrm{i}^\prime a}$ (see \p{eta})
we obtain from \p{J5-st-a} the value of the Casimir operator \p{J5-st}:
\begin{equation}
\label{J5-st-ab}
\hat C_6 \ = \ -  \mu^2 \, J_{\,\mathrm{i}}J_{\,\mathrm{i}}\,,
\end{equation}
where
\begin{equation}
\label{Ji-st}
J_{\,\mathrm{i}}\ := \ M^{(+)}_{\,\mathrm{i}} +M^{(-)}_{\,\mathrm{i}}
\end{equation}
are the generators of the diagonal ${su}(2)$ subalgebra of the ${so}(4)={su}(2)\oplus{su}(2)$ stability algebra.
Using \p{Mpm} and \p{Mpm-eta} and explicit expressions of the ‘t Hooft symbols  (see e.g. Sect. 3.3.3 in
\cite{IR})
we find
\begin{equation}
\label{Ji-st-ex}
J_{\,\mathrm{i}}\ = \ -\frac12\, \epsilon_{\mathrm{i}\mathrm{j}\mathrm{k}} M_{\,\mathrm{j}\mathrm{k}} \,, \qquad \mathrm{i}=1,2,3\,.
\end{equation}
So the operators \p{Ji-st} are in fact the generators of the $\mathrm{SO}(3)$ subgroup of the $\mathrm{SO}(4)$ stability group.
Therefore,  in case of the unitary representations it is necessary to satisfy the equality
\begin{equation}
\label{MM-st}
J^2 \ = \ s(s+1)\,,
\end{equation}
where $s$ is fixed integer or half-integer number.

So, in case of the irreducible representations of infinite
(continuous) spin, the Casimir operator \p{J5-2-Cas} takes the value
\begin{equation}
\label{J5-st-abc}
C_6 \ = \ \hat C_6 \ = \ - \mu^2 \, s(s+1) \,,
\end{equation}
Such irreducible representations describe a tower of infinite number
of massless states.

As a result, the massless infinite spin representations are
characterized by the pair $(\mu,s)$, where the real parameter $\mu$
defines the eigenvalue of the Casimir operator \p{W-2-n0} and the
(half-)integer number $s$ defines the eigenvalue of the Casimir
operator \p{J5-st-abc}.

Let us examine in our consideration the $D=6$ infinite integer spin system \cite{BekMou}
which is higher dimension generalization of the $D=4$ model \cite{Wigner39}, \cite{Wigner47}, \cite{BargWigner}.
This model \cite{BekMou} is described by
the pair of the space-time phase operators
\begin{equation}
\label{x-p}
x^m\,, p_m\,,\quad [x^m,  p_k]=i\delta^m_k
\end{equation}
and two pairs of the additional bosonic phase vectors
\begin{equation}
\label{y-q}
w^m\,, \xi_m\,,\quad [w^m,  \xi_k]=i\delta^m_k\,;\qquad\quad u^m\,, \zeta_m\,,\quad [u^m,  \zeta_k]=i\delta^m_k\,.
\end{equation}
These two pairs of vectors \p{y-q}
 are responsible for spinning degrees of freedom.

Infinite integer spin field $\Psi$ in  \cite{BekMou} is described by the $D=6$ generalization of the Wigner-Bargmann equations
\begin{eqnarray}
\label{P2-equ}
p^2\,\Psi &=&  0\,, \\ [7pt]
\label{WB1}
\xi\!\cdot\!p\,\Psi&=&  0\,, \\ [7pt]
\label{WB2}
(w\!\cdot\!p-\mu)\,\Psi&=&0\,, \\ [7pt]
\label{WB3}
(\xi\!\cdot\!\xi+1)\,\Psi&=&0\,,
\end{eqnarray}
and additional equations with vectorial operators from the second pair \p{y-q}
\begin{eqnarray}
\label{add-equ-1}
u\!\cdot\!p\,\Psi &=&  0\,, \\ [7pt]
\label{add-equ-2}
\zeta\!\cdot\!p\,\Psi &=&  0\,, \\ [7pt]
\label{add-equ-3}
\zeta\!\cdot\!\xi\,\Psi&=&  0\,, \\ [7pt]
\label{add-equ-4}
\zeta\!\cdot\!\zeta\,\Psi&=&0\,, \\ [7pt]
\label{add-equ-5}
(u\!\cdot\!\zeta-s)\,\Psi&=&0\,,
\end{eqnarray}
where $\xi\!\cdot\!p:=\xi^m p_m$, \textit{etc}.

Note that, in contrast to the four-dimensional case  \cite{Wigner39}, \cite{Wigner47}, \cite{BargWigner}
with one pair of auxiliary variables $w^m$, $\xi_m$,\footnote{Note that in the twistor formulation of the infinite spin particle \cite{BFI},
it was more convenient for us to use dimensional additional variables $y^m=w^m/\mu$, $q_m=\mu \xi_m$.}
in the six-dimensional case it is necessary to use the second pair of auxiliary vector variables
$u^m$, $\zeta_m$ to describe arbitrary infinite spin representations.

In the light-cone frame \p{P-st1}, \textit{i.e.} $p^-=p_a=0$, $p^+=\mbox{const}\neq0$,
and in the representation $\xi_m=-i\partial/\partial w^m$, $\zeta_m=-i\partial/\partial u^m$ the equations \p{WB1}-\p{WB3} give the conditions
\begin{eqnarray}
\label{WB1-lc}
\frac{\partial}{\partial w^+}\,\Psi&=&  0\,, \\ [7pt]
\label{WB2-lc}
(p^+w^- - \mu)\,\Psi&=&0\,, \\ [7pt]
\label{WB3-lc}
\left(\frac{\partial}{\partial w_a}\frac{\partial}{\partial w_a}+1\right) \Psi&=&0\,,
\end{eqnarray}
whereas \p{add-equ-1}-\p{add-equ-5} yield
\begin{eqnarray}
\label{add-lc-1}
p^+u^-\,\Psi&=&  0\,, \\ [7pt]
\label{add-lc-2}
\frac{\partial}{\partial u^+}\,\Psi&=&  0\,, \\ [7pt]
\label{add-lc-3}
\frac{\partial}{\partial u_a}\frac{\partial}{\partial w_a}\,\Psi&=&  0\,, \\ [7pt]
\label{add-lc-4}
\frac{\partial}{\partial u_a}\frac{\partial}{\partial u_a}\,\Psi&=&0\,, \\ [7pt]
\label{add-lc-5}
\left(u_a\frac{\partial}{\partial u_a}-s\right) \Psi&=&0\,,
\end{eqnarray}
The solution of the equations \p{WB1-lc}-\p{add-lc-5} is the field
\begin{equation}
\label{Psi-expres}
\Psi = \delta(p^+w^- - \mu)\,\delta(p^+u^-)\,\Phi(w_a,u_a)\,,
\end{equation}
where $\Phi(w_a,u_a)$ is subjected \p{WB3-lc}, \p{add-lc-3}-\p{add-lc-5} and has series expansions presented in  \cite{BekMou}.

Now we can determine the values of the Casimir operators  \p{W2-st}, \p{J5-st} on the field \p{Psi-expres}.

For the field \p{Psi-expres} the generators of the $iso(4)$ algebra \p{P-J}, \p{J-J}
have the form
\begin{equation}
\label{P-J-lc}
M_{ab}=i\left(w_a\, \frac{\partial}{\partial w_b} -w_b\, \frac{\partial}{\partial w_a}
+u_a\, \frac{\partial}{\partial u_b} -u_b\, \frac{\partial}{\partial u_a}\right)  \,,\qquad
\hat\Pi_{a}= -i\mu \frac{\partial}{\partial w_a}\,.
\end{equation}
As result, due to the equation \p{WB3-lc}, we obtain the fulfillment of the condition \p{W-2-n0} for the Casimir operator $C_4$:
$C_4=\hat C_4=-\mu^2$.
Moreover, the representations \p{P-J-lc} lead to the expression
\begin{eqnarray}
\label{P-2-vec}
\hat C_6
&=&
\mu^2\, u_a\frac{\partial}{\partial u_a}\left(u_b\frac{\partial}{\partial u_b}+1 \right)
\frac{\partial}{\partial w_c} \frac{\partial}{\partial w_c}
\\ [7pt]
\nonumber
&&
+\,\mu^2\left(u_a\frac{\partial}{\partial w_a}\,u_b\frac{\partial}{\partial w_b} \ - \
u_a u_a \,\frac{\partial}{\partial w_b} \frac{\partial}{\partial w_b}\right)\frac{\partial}{\partial u_c} \frac{\partial}{\partial u_c}
\\ [7pt]
\nonumber
&&
+\,\mu^2\left(u_a u_a \,\frac{\partial}{\partial u_b} \frac{\partial}{\partial w_b}
 \ - \
2u_a\frac{\partial}{\partial u_a}\,u_b\frac{\partial}{\partial w_b}\right)\frac{\partial}{\partial u_c} \frac{\partial}{\partial w_c}
\end{eqnarray}
for the sixth order Casimir operator.
So, due to the equations \p{WB3-lc}, \p{add-lc-3}-\p{add-lc-5} the operator \p{J5-st} takes the value  $C_6=\hat C_6=-\mu^2 s(s+1)$ on the field  \p{Psi-expres}.

Thus, the infinite spin field with only one additional vector variables and
obeying the Wigner-Bargmann equations \p{P2-equ}-\p{WB3} and additional equations  \p{WB1-lc}-\p{add-lc-5}
describes the irreducible  $(\mu,s)$ infinite spin representation.
The system with only one pair of auxiliary variables $w^m$, $\xi_m$ in \p{y-q}
(without using the second pair of auxiliary vector variables $u^m$, $\zeta_m$)
and with only the equations of motion \p{WB1}-\p{WB3} describe the infinite spin representations at $s=0$ \cite{BekMou}.

\setcounter{equation}0
\section{Summary and outlook}

We have studied the massless irreducible representations of the
Poincar\'{e} group in six-dimensional Minkowski space. The representations are described by
three Casimir operators written in the form (\ref{2-Cas-f}),
(\ref{4-Cas-f}), (\ref{J5-2-Cas-f}) or in the equivalent form
(\ref{P-2-Cas}), (\ref{W3-2-Cas}), (\ref{J5-2-Cas}). The properties
of these operators are explored in the standard massless momentum
reference frame, where it is seen that the unitary representations
of $ISO(1,5)$ group are induced from representations of $SO(4)$ and
$ISO(4)$ groups and correspondingly are divided into finite spin
(helicity) and infinite spin representations. Both these
representations are studied in details. It is proved that the finite
spin representation is described by two integer or half-integer
numbers while the infinite spin representation is described by one
real parameter and one integer or half-integer number. In
case of half-integer spin we should introduce an additional spinor
or twistor variables like in \cite{BekMou}.
The results obtained here are in agreement with the statements of
\cite{BKRX,BekBoul} where it was done the classification of massless representations
of any dimension Poincar\'{e} group via the induced representation method.

As a continuation of this research it would be interesting to
describe the massless representations with half-integer spin and
massive irreducible representations of six-dimensional Poincar\'{e}
group with both integer and half-integer spin. Another open problem
is constructing the representations of the corresponding
six-dimensional {\em super} Poincar\'{e} group. Also it would be
useful to work out the field realizations of the massless
representations considered in this paper (see, e.g., Remark at the
end of Sect.\,\ref{3nrt}) and explore the new
aspects of Lagrange formulation for these fields in six-dimensional
Minkowski space including infinite spin
cases.\footnote{Currently, there is a fairly large
literature on various aspects of infinite spin Lagrange formulation
(see e.g., the recent paper \cite{KZin} and the references
therein).}
We plan to study all these problems in the forthcoming papers.

\section*{Acknowledgments}

We  are grateful to
N. Boulanger, S.M. Kuzenko  and  Yu.M. Zinoviev  for
correspondence. Work of I.L.B, S.A.F and A.P.I was supported in
part by The Ministry of Education of Russian Federation,
project FEWF-2020-0003. I.L.B is grateful to RFBR grant, project No
18-02-00153. M.A.P is grateful to RFBR grant, project No
19-01-00726А.

\bigskip

\renewcommand\theequation{A.\arabic{equation}} \setcounter{equation}0
\section*{Appendix.  On the Casimir operators of the Poincar\'e algebra}

The quantity $\varepsilon^{mnklpr}W_{mnk}W_{lpr}$ could be an additional Casimir operator for $\mathfrak{iso}(1,5)$ algebra.
But it is identically equal to zero.
This fact is a special case of the property of any rank $r$ antisymmetric tensor
$W_{m_1\ldots m_r}$ in $2r$-dimensional space, when $r$ is odd number.
Indeed, in this case we have $(W,V)_\varepsilon=(-1)^r(V,W)_\varepsilon$,
where $(W,V)_\varepsilon :=\varepsilon^{m_1\ldots m_rn_1\ldots n_r}W_{m_1\ldots m_r}V_{n_1\ldots n_r}$
and $\varepsilon^{m_1\ldots m_rn_1\ldots n_r}[W_{m_1\ldots m_r},V_{n_1\ldots n_r}]=0$.
Thus, for antisymmetric tensor with components
$$
W_{m_1\ldots m_r}=\varepsilon_{m_1\ldots m_rn_1\ldots n_r}P^{n_1}M^{n_2n_3}\ldots M^{n_{r-1}n_r} \; ,
$$
which is defined only for odd $r$, we always have $(W,W)_\varepsilon=0$.
In this case
 a Casimir operator for $\mathfrak{iso}(1,2r-1)$ algebra, of the second order in $W$,
has the unique form
$$
W^2=\frac{1}{(r+1)!}\,W^{m_1\ldots m_r}W_{m_1\ldots m_r}\,.
$$
Whereas for even $r$ we have antisymmetric tensor with components
$$
L_{m_1\ldots m_r}=\varepsilon_{m_1\ldots m_rn_1\ldots n_r}M^{n_1n_2}\ldots M^{n_{r-1}n_r}
$$
which yields for $\mathfrak{so}(\ell,2r -\ell)$ algebra
additional to $L^2=L^{m_1\ldots m_r}L_{m_1\ldots m_r}$
Casimir operator $(L,L)_\varepsilon
\neq 0$
(see  operator \p{Lambda-st-01} written for
 the case of $\mathfrak{so}(4)$ algebra).

\end{document}